# Quantifying AI impact in energy transitions: The Energy Justice Impact Assessment (EJIA) framework

Emily Bringmann[a], Florian Kutzner[a], Bianca Weber[a], and Celina Kacperski[a,b]

[a]Seeburg Castle University, Seeburgstraße 8, 5201 Seekirchen am Wallersee, Austria

[b]University of Konstanz, Universitätsstraße 10, 78464 Konstanz, Germany

emily.bringmann@uni-seeburg.at

florian.kutzner@uni-seeburg.at,

celina.kacperski@uni-seeburg.at

Corresponding author:

Celina Kacperski, celina.kacperski@uni-seeburg.at

**Abstract**

Artificial intelligence is increasingly deployed across energy systems to optimize efficiency, balance supply and demand, and integrate renewable sources. These applications alter how energy systems function, changing how benefits and burdens are distributed, whose needs are represented in system design, and who can influence decision-making, with consequences for energy justice that are rarely quantified. This paper addresses that gap through a systematic review of AI impact assessment approaches, followed by the development of a new framework. Reviewing 26 peer-reviewed frameworks using PRISMA methodology, we find that most address a single impact dimension, only three provide quantifiable indicators, and although 21 reference justice implications, only two operationalize justice-relevant constructs through measurable indicators. No existing framework combines multidimensional impact coverage with quantifiable indicators disaggregated by stakeholder group. We therefore introduce the Energy Justice Impact Assessment (EJIA) framework. Crossing four AI lifecycle stages – data collection, model development, deployment, and continuous adaptation – with the three tenets of energy justice (distributional, recognition, procedural) produces a map of where injustice can arise, which we use to derive a set of environmental, social, and economic outcome indicators. These are measured separately per affected stakeholder group and assessed against a counterfactual across defined phases, enabling observed effects to be attributed to the AI application. The EJIA framework is descriptive rather than normative: it makes differential impacts visible and comparable, while judging whether a pattern constitutes injustice remains a context-dependent decision for practitioners. An exemplary application to an AI-based demand-side response system in social housing demonstrates the framework's use.

Word count: 7929

## 1. Introduction

The European "just energy transition" aims at transitioning the European energy system from fossil to renewable energies without structural disadvantages for vulnerable groups [1]. In the energy transition, artificial intelligence (AI) has huge potential for optimizing efficiency, harmonizing supply and demand, and for improved predictive maintenance of energy production sites [2]. Also, AI technologies are relatively cost-efficient solutions for integrating renewable energy sources into the energy system in comparison to electricity storage [3]. At the same time, AI has documented negative impacts, such as a high energy demand and displacement of human labor expertise despite unclear gains [4,5]. It is evident that AI solutions applied in the energy transition are beginning to structurally alter how energy systems function, introducing risks and chances for energy justice.

AI's impact is often discussed in terms of its biases [6]. Since algorithm outcomes depend on the data they are trained on, biased outcomes can occur when training data largely over- or underrepresents different stakeholder groups. However, AI can impact energy justice beyond data: for example, it changes the core system logic (e.g., pricing, allocation, forecasting)[7,8], it is based on algorithms that are often invisible or non-transparent, making downstream harm harder to detect and contest [7], and it shifts control over energy system operation increasingly toward actors (e.g., private technology firms) whose commercial priorities have not traditionally been oriented toward energy justice [8]. This can lead to shifts in how energy-related benefits and burdens are distributed, which groups and needs are represented in system design, and who is able to participate in or influence decision-making processes.

Through these mechanisms, AI offers both potential for improvements of energy justice as well as risks of impediments. Because of its rapid rollout, impacts on energy justice are rarely quantified, and if they are, results are not necessarily comparable, because different measurement methodologies are applied. Many scholars and organizations have been stressing the need to quantify these diverse effects of AI in energy [e.g., 6,7,9]. Impact assessment is also becoming a legal requirement: the EU AI Act obliges deployers of high-risk AI systems to assess impacts on fundamental rights such as non-discrimination before deployment [10]. This paper proposes the Energy Justice Impact Assessment (EJIA) framework for AI that encompasses multidimensional impacts and their implications for energy justice. It builds on existing work showing how algorithmic bias arises at different stages of the AI lifecycle and relates to energy justice [6], and extends this diagnostic account beyond algorithmic bias, deriving quantifiable indicators disaggregated by stakeholder group, and assessed at defined phases.

## 2. Theoretical background

### 2.1 Applications of AI in energy

Energy is one of the most prominent fields for machine learning applications, accounting for 46.6% of subject-area usage – more than engineering (23.1%) and environmental science (11.2%) combined [11]. Applications of AI in energy take place both on the production as well as on the consumption side. Common applications are the optimization of heating ventilation and air conditioning systems [12], consumption and price forecasting and optimization [13], and predictive maintenance of energy plants such as solar parks [2]. Optimally, AI use then results in enhanced energy efficiency and grid stability, improved integration of renewable energy, and reduced operational costs [2]. As downstream consequence, AI can thus reduce greenhouse gas emissions and thereby contribute to or even enable the energy transition with the goal of setting the base for a net zero emissions Europe [14].

## 2.2 Energy justice and AI

In their seminal work, McCauley et al. [15] distinguish between distributional, recognition, and procedural dimensions of energy justice. Distributional justice is concerned with the distribution of benefits and costs caused by the energy system and topics of interest have traditionally been the siting of energy infrastructure or equal access to energy services [15,16]. Recognition justice looks at minorities or vulnerable populations and the (mis-)representation of their needs in a system. A prominent example of a recognition justice issue is the debate over local opposition to wind farms: recognition justice scholars argue that framing protesters merely as self-interested or misinformed can overlook the values, identities, and concerns that motivate opposition, as well as demands for fair and respectful treatment [15,16]. Finally, procedural justice revolves around the inclusion of relevant stakeholders in decision-making processes, and transparency of processes. Measures such as active inclusion of local knowledge, public information disclosure, and institutional representation of affected groups have been topics of interest [15,16].

The integration of AI in the energy system comes with new issues for energy justice [7]. Besides the algorithms themselves along with their inevitable potential biases, their application context as well as the power structures behind them can positively or negatively affect energy justice. Distributional justice is invoked as economic costs and profits are redistributed, including when public funding, knowledge sharing, and data governance are reconfigured as AI is introduced into energy systems. Recognition justice becomes relevant in the selection of communities and participants for smart grid projects, as well as in questions concerning the accessibility of smart technologies and infrastructure for different user groups – both economically and in terms of digital or technical literacy. Procedural justice concerns the participation of different stakeholders in decision-making processes around AI-based energy systems. It also raises questions of user control in relation to automated systems, as well as transparency in the use and handling of data collected in smart grid projects. In a case study at the regional level, Ye et al. [17] find that AI adoption in China has been associated with worsening distributional justice, at least initially, while simultaneously improving procedural and recognition justice. They proxy energy justice through 24 indicators drawn from national statistical yearbooks such as insurance coverage rates, electricity prices, investment ratios, or dependency ratios which are aggregate values at regional level.

Both costs and benefits for different stakeholder groups have rarely been empirically quantified and are mostly collections of potential effects [6,18,19], at most organized along the three justice tenets. Where Ye et al. [17] do quantify real-world impact, results remain at regional level rather than at the level of specific stakeholder groups or concrete AI applications. In the following, we will provide two examples for potential impacts of AI on energy justice to stress the need for coherent quantification.

#### 2.2.1 AI-powered demand-side response as a justice-relevant innovation.

Demand-side response (DSR) refers to the adjustment of electricity consumption by end users in response to price signals, incentives, or grid requirements, and is widely discussed as a key mechanism for increasing flexibility in smart grids [20]. AI and machine learning can support tasks such as identifying suitable consumers for demand response, learning user preferences and consumption patterns, designing dynamic pricing schemes, scheduling and controlling flexible devices through forecasting of consumption and production, and determining incentives or rewards for participation [21]. Thus, AI has become increasingly relevant for DSR because it involves complex, data-intensive, and often near real-time decisions.

At the same time, the growing use of DSR raises important questions regarding the distribution of its costs and benefits. Flexible tariffs linked to DSR are often perceived as more just than fixed tariffs by both grid representatives and households [e.g., 22]. Unlike fixed tariffs, they allow consumers to reduce costs not only by using less electricity but also by shifting consumption to periods of lower prices [20]. Computational models suggest that even modest load shifting can generate meaningful savings, although benefits vary across households and require a certain minimum degree of flexibility [23,24]. These findings indicate that DSR can support distributional justice goals if its benefits are broadly accessible. However, households structurally vary in their capability to behave flexibly, and thereby to gain from flexible energy prices. Powells and Fell [25] introduce the concept of energy flexibility capital as the capacity of consumers to adapt their energy usage patterns in response to system needs. They claim that its source varies systematically with affluence: less affluent households' flexibility capital tends to be socially derived, stemming from changes to daily habits and routines, while more affluent households' flexibility capital tends to be technologically derived, through devices such as batteries and automated or smart appliances [25]. This means less affluent households are often required to adapt their habits and daily routines to benefit from flexible pricing, even though they may lack the capacity to do so. White and Sintov [26] found in a large-scale randomized controlled trial (n = 7,487 households) that assigning households to time-of-use pricing increased monthly bills by an additional 7-9 dollars for elderly- and disability-vulnerable households relative to non-vulnerable households, and predicted worse health outcomes (e.g., likelihood of needing medical attention for heat-related reasons) for households with disabled or Hispanic occupants. These unequally distributed outcomes violate distributional justice principles, while the lack of consideration of different stakeholder groups' needs and capabilities threatens recognition justice.

Issues of distributional and recognition justice also arise because households' ability to provide energy flexibility is shaped by factors such as housing insulation [5], digital literacy [25], and climate and weather conditions [27].

### 2.2.2 AI-supported grid planning and infrastructure investment optimization

AI algorithms can be used to identify siting areas for energy infrastructure such as solar or wind parks, and for prioritizing investment areas [18,28], e.g. for smart meter rollout or DSR pilot sites. This could help identify energy-poor or grid-vulnerable areas for investment when doing so also improves revenue or grid stability [28]. As a result, AI could contribute to distributional justice by expanding access to affordable and reliable energy in underserved communities [28]. If explicitly incorporated into algorithm design, indicators such as energy poverty, outage vulnerability, or poor infrastructure quality can also support recognition justice by ensuring that the needs of different social groups are considered [19,28]. Procedural justice may likewise benefit if AI systems make the criteria underlying decisions transparent [19].

The actual justice outcome of the usage of these AI tools largely depends on the input data. Underserved, infrastructure-poor areas with lower-income households are likely to produce fewer data points that an AI system can be trained on, which may skew AI models to prefer better-resourced households and disadvantage lower-income ones [6]. Beyond input data, the optimization objectives themselves can also reproduce distributional inequalities. If these objectives are defined around economic efficiency, revenue maximization, or property values, the model may indirectly prioritize socioeconomically advantaged areas, as these criteria often correlate with affluence. This risk is even more direct when demographic indicators are used to estimate investment value, potentially steering infrastructure investments toward already affluent areas while underserved, energy-poor, or grid-vulnerable communities remain deprioritized [6]. Recognition justice is at risk when the indicators guiding AI-supported planning fail to capture the specific needs and constraints of vulnerable groups. Even where underserved areas are included as possible investment targets, the model may overlook why these communities are vulnerable, for instance due to unreliable grid access, digital literacy, or disability [19,28].

Procedural justice may also be affected if AI-supported planning decisions are not sufficiently transparent. When investment priorities are produced by complex or opaque models, affected communities may have little insight into why some areas are prioritized over others, which indicators were considered, or how competing goals such as efficiency, profitability, and equity were weighted. This makes it harder to question or challenge decisions and may limit meaningful participation in infrastructure planning [19,28].

To our knowledge, apart from White and Sintov's [26] randomized controlled trial on distributional economic outcomes, no quantitative empirical study has examined the justice-relevant risks and benefits of AI in energy systems at the level of specific stakeholder groups. This lack of empirical evidence underscores the need for a systematic impact assessment framework capable of identifying and quantifying justice-related risks and benefits across different stakeholder groups [6,7,18].

### 2.3 Impact assessment of AI across domains

Employing PRISMA, below, we review impact assessment frameworks for AI. To give a brief introduction here, we find that in general, frameworks tend to be distinguished based on their primary focus. Some approaches are strongly compliance-oriented, aiming to align AI systems with regulatory requirements and reporting standards [e.g., 29,30]. Bogucka et al. [29], for example, present a ready-to-use template for reporting alignment with AI regulations such as compliance and ethical responsibilities, requiring practitioners to document e.g. risks and benefits for fundamental rights, safety, and democracy. A comparable logic underlies institutional standardization: ISO/IEC 42005:2025 provides a taxonomy of AI's potential harms and benefits and an example assessment template, leaving their operationalization into concrete indicators to the implementing organization [31]. Other frameworks are purely risk-oriented, focusing on identifying potential harms associated with AI such as cyberattacks, bioweapon development, and disinformation [32], or broader harm categories such as discrimination, privacy and security violations, or misinformation: Slattery et al. [33] have compiled a risk repository, distinguishing between different causal factors of AI risks. A third group centers on specific normative domains such as human rights or ethical principles and how to incorporate these into AI system design [e.g., 34–36]. Mantelero and Esposito [34], for example, present a human rights impact assessment approach, allowing the classification of risks to different facets of human rights such as human dignity and freedom from discrimination, and according to their likelihood and severity before their real-world implementation.

Finally, Stahl et al. [37] provide a cross-domain systematic review of AI impact assessment approaches and identify substantial heterogeneity in scope, purpose, and implementation. They highlight the fragmentation of the field and the prevalence of narrowly scoped approaches focused on single dimensions such as privacy, environmental effects, or ethics. They do not identify an impact assessment framework for AI in the energy context focusing on energy justice outcomes. Therefore, in the following we build on their mapping and description of existing frameworks, and we offer a structured cross-framework comparison of how specific impact dimensions are operationalized, and systematically evaluate frameworks with respect to their suitability for the application domains such as energy systems.

In summary, the present systematic review updates and extends the evidence base established by Stahl et al. [37] by including more recent frameworks identified through broader search terms including publications after 2021. In addition, it applies a shared, energy-justice-based coding framework to systematically compare AI impact assessment approaches across studies. This enables a directly comparable analysis of whether and how existing frameworks incorporate distributional, recognition, and procedural justice considerations, whether they translate these into quantifiable indicators rather than offering purely process-oriented guidance, and whether they differentiate impacts across distinct stakeholder groups. The resulting cross-framework comparison thereby allows the evaluation of the frameworks' applicability to assess justice impacts of AI in energy systems.

## 3. Systematic literature review

The goal of this section is to present a systematic literature review of impact assessments of AI, categorizing studies as to the type of AI, and impact in focus, the inclusion of justice implications, and if measurable indicators are present. We first outline our search strategy and analysis criteria. We then present our results for cross-domain frameworks, and the implications for our goal of creating a justice-oriented multi-dimensional impact assessment of AI in energy.

### 3.1 PRISMA methodology

We applied a systematic search strategy in line with the PRISMA guidelines [38] in Web of Science, ScienceDirect, and IEEE using the following search strings (see Figure 1):

- "impact evaluation" AND (AI OR "artificial intelligence") NOT (LLM OR "generative artificial intelligence" OR GenAI)

- “impact assessment” AND (AI OR “artificial intelligence”) NOT (LLM OR “generative artificial intelligence” OR GenAI)
- “evaluat*” AND framework AND (AI OR “artificial intelligence”) NOT (LLM OR “generative artificial intelligence” OR GenAI)
- “assess*” AND framework AND (AI OR “artificial intelligence”) NOT (LLM OR “generative artificial intelligence” OR GenAI)

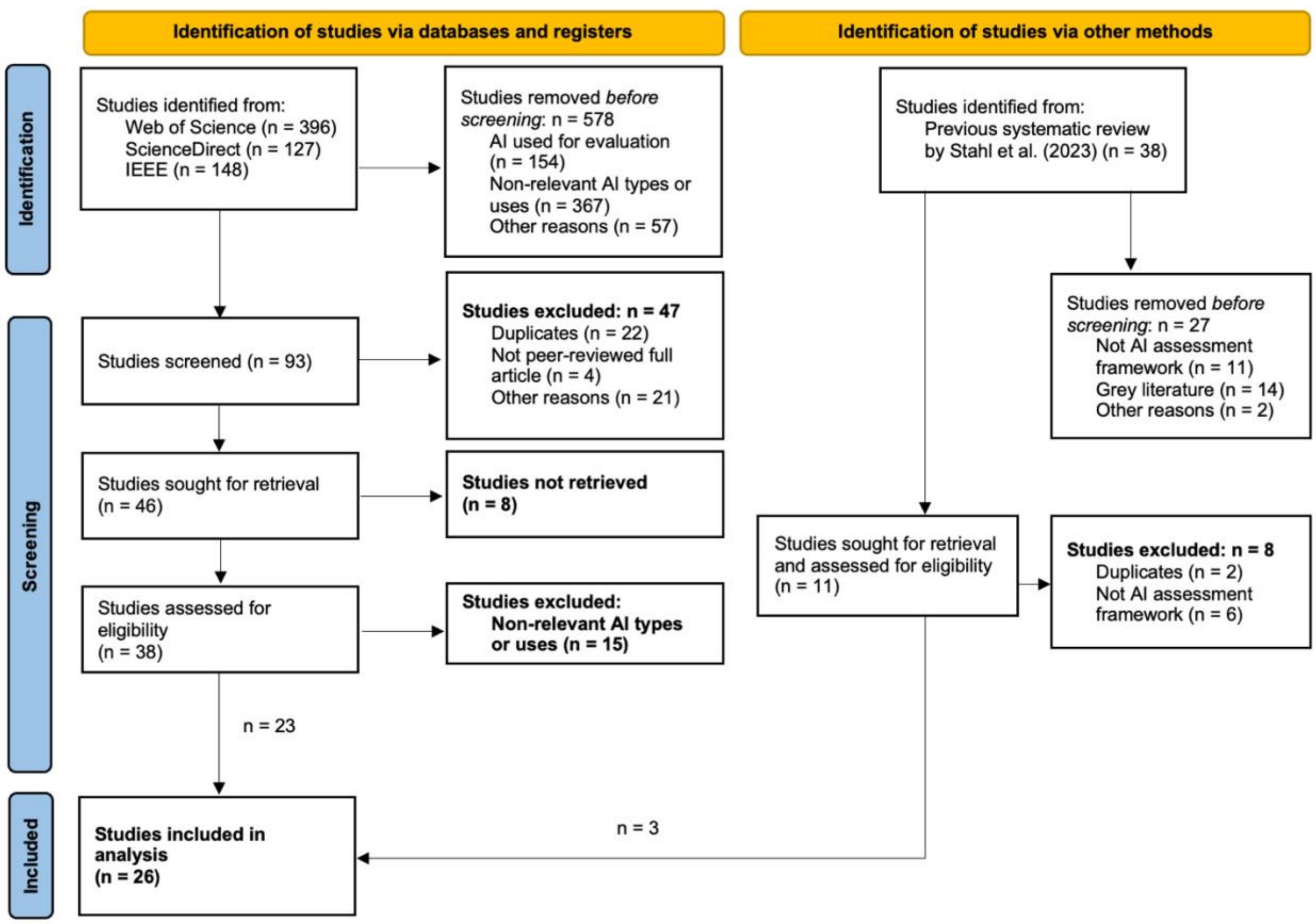

**Figure 1.** Flowchart of systematic literature review for identifying relevant articles on impact assessment approaches of AI based on PRISMA [38]. Studies out-scoped due to “other reasons” e.g. did not present replicable assessment methodology.

Given the focus of this review on evaluating AI application frameworks for suitability in energy systems, studies centered exclusively on large language models (LLMs) were excluded, as these typically address application domains and evaluation criteria that are not directly transferable to energy system contexts. This led to a total of 671 identified studies published until March 2026. As only studies that might be applicable to the context of energy were to be evaluated, we also excluded studies on impact assessment in a specific, non-relevant field such as medicine or finance, as well as studies focusing on impacts within an organization only. Articles presenting case-specific results of conducted impact assessments were excluded if they do not include a generalizable methodology. Only peer-reviewed articles available in English were included. Additionally, peer-reviewed studies from Stahl et al.'s [37] systematic review were reviewed and included if they were eligible and not already included via our own search terms. The overlap with Stahl et al. [37] proved minimal, with only N=5 studies in both reviews (two identified through our literature research).

### 3.2 Results of systematic literature review

Table 1 depicts the final 26 studies, which we analyzed for AI type, impact focus, inclusion of justice implications, and presence of measurable indicators. Generally, we found that most frameworks focused exclusively on one impact dimension such as trustworthiness [35,36,39] or human rights [34,40,41]. Only two studies [36,42] focused on AI applications noting impacts directly related to energy outcomes, but neither was a multi-dimensional framework or included concrete indicators for their impact dimension(s) in focus. Assessment methodologies that focused on multiple impact dimensions did not include quantifiable indicators but were solely process-oriented, without complementary outcome indicators that could capture distributive effects [29,43–47]. This lack of presentation of quantifiable indicators was a general trend throughout the reviewed publications. Only von Zahn et al. [48], Bertaina et al. [40] and Borraccia et al. [49] included quantifiable impact indicators, while four frameworks included a checklist of indicators [50–53]. All other reviewed publications offered either solely process-oriented frameworks that aim at guiding the user in designing use case specific indicators, or they were high-level or of conceptual nature.

Justice considerations played a role in a substantial number of frameworks (21 out of 26 publications include justice implications). Although this hints to the scholars' understanding of the relevance of justice aspects in the context of AI, only de Fine Licht and Folland's [54] and Bertaina et al.'s [40] frameworks allowed to structurally examine the differential impact of AI on different stakeholder groups. Many frameworks claimed that distributional, recognition, and/or procedural justice impacts could be assessed through their framework, however, they did not present any concrete methodology such as quantifiable indicators for doing so. Only two frameworks in this review offered quantifiable indicators for justice-relevant constructs [40,48]. All other publications mentioning justice or corresponding constructs [in line with 15, and 16] referred to them in loose, qualitative terms rather than through a concrete, quantifiable methodology. Finally, the majority of the frameworks focused on the assessment of negative impacts (19 of 26), often in the form of risks, not on a comprehensive impact mapping.

These findings point to a clear gap: To date, there is no multidimensional impact assessment framework of AI in energy including quantifiable indicators that can be mapped onto different impacted stakeholder groups.

**Table 1**

Overview of 26 reviewed AI impact assessment frameworks.

| Year | Authors | Type of AI | Impact in focus | Justice implications | Justice tenet(s) | Valence of impacts addressed | Quantifiable impact indicators |
|---|---|---|---|---|---|---|---|
| 2025 | Bertaina et al. [40] | All types of AI | fundamental rights: 50 fundamental rights by EU. | Yes | Distributional, recognition and procedural justice | Negative | Yes |
| 2024 | Bogucka et al. [29] | All types of AI (with focus on machine learning approaches) | Risks, mitigation strategies, and benefits of AI's (1) capabilities, (2) human interaction, and (3) systematic impact. | No | - | Positive and negative | No; process-oriented, assists in identifying impacts |
| 2025 | Bogucka et al. [43] | All types of AI | Benefits and capability risks, human interaction risks, and systemic risks for the three stakeholders AI subject, AI deployer, and institutions and environment; performance of applied models on data | Yes | Not further specified | Positive and negative | No; process-oriented |
| 2025 | Borraccia [49] | All types of AI | Environmental impact in kgCO2e (kilograms of CO2-equivalent emissions) (costs only) | No | - | Negative | Yes |
| 2025 | de Fine Licht and Folland [54] | AI applied in public sector decision-making | well-being (as a means to identify harms and benefits); trustworthiness: performance, explainability, fairness, legality, and accountability<br><br>they include stratifying impact analysis by demographic groups to identify disparate effects | Yes | Distributional and procedural justice | Positive and negative | Neither: framework largely process-based but includes concrete operationalizations e.g. well-being questionnaires to be applied. |
| 2026 | Eke et al. [50] | Technologies in general, particularly (all types of) AI | colonial structure in technologies: inherent colonial legacies in technologies such as AI, global power asymmetries, and epistemic injustices | Yes | Distributional, recognition and procedural justice | Negative | No; checklist format |
| 2024 | Fontes et al. [44] | Urban AI (AI systems for public service and urban governance) | Collective societal challenges of AI in urban governance (data/privacy dependency, algorithmic bias, impacts on human autonomy, AI literacy, and AI sustainability) | Yes | Procedural and recognition justice | Negative | No; process-oriented |
| 2024 | Goldstein and Sastry [32] | Advanced machine learning systems | Risk of malicious use of AI according to their plausibility, performance and observed use | No | - | Negative | No; process-oriented |
| 2023 | Havrda and Klocek [55] | All types of AI | well-being as a multidimensional outcome lens influenced by AI impacts | No | - | Positive and negative | No; process-oriented |
| 2020 | Ivanova [30] | AI based on algorithms with the potential to lead to discrimination | Data protection and non-discrimination | Yes | Distributional and recognition justice | Negative | No; process-oriented |
| 2024 | Korobenko et al. [51] | All types of AI | Privacy, cybersecurity, and organizational governance assessed on the four dimensions | Yes | Procedural justice | Negative | No; checklist format |
| | | | data, technology, people, and process | | | | |
| 2023 | Landers and Behrend [52] | High-complexity predictive models used to make predictions about individual cases | Fairness and bias | Yes | Distributional and procedural justice | Negative | No; checklist format |
| 2018 | Mantelero [56] | Data-intensive AI technologies | Data protection, fundamental rights, and freedoms (e.g. of movement, expression, or freedom in the workplace) discrimination, social justice, societal well-being, fairness and accountability | Yes | Recognition and procedural justice | Negative | No; conceptual blueprint |
| 2024 | Mantelero [41] | All types of AI | Fundamental rights | Yes | Not further specified | Negative | No; process-oriented |
| 2021 | Mantelero and Esposito [34] | Data-intensive AI technologies | Risks to human rights (fundamental rights and freedom) on both individual and collective level, e.g., privacy and data protection | Yes | Not further specified | Negative | No; process-oriented |
| 2022 | Nitta et al. [35] | All types of AI | 7 key factors of trustworthiness[1] | Yes | Distributional, recognition and procedural justice | Negative | No; process-oriented |
| 2024 | Pelekis et al. [36] | AI in energy | 7 key factors of trustworthiness1 | Yes | Distributional, recognition and procedural justice | Negative | No; process-oriented |

[1] human agency and oversight; technical robustness and safety; privacy and data governance; transparency; diversity, nondiscrimination, and fairness; societal and environmental well-being; accountability.

| | | | | | | | |
|---|---|---|---|---|---|---|---|
| 2025 | Pudney et al. [45] | AI in critical infrastructure | Service fulfillment, health and safety, economic effectiveness, environmental sustainability, socioeconomic well-being, privacy and security, governance and compliance | Yes | Distributional, recognition and procedural justice | Positive and negative | No, only exemplarily; process-oriented |
| 2020 | Raji et al. [57] | All types of AI | Ethical and social impact, non-discrimination | Yes | Distributional, recognition and procedural justice | Negative | No; process-oriented |
| 2021 | Sætra [46] | All types of AI | Impact on the SDG goals: environmental, equality-, economic growth-, health-, and work-related goals. | Yes | Not further specified | Positive and negative | No; present methodology, concrete indicators not yet developed |
| 2020 | Schiff et al. [58] | All types of AI | Impact on human and societal well-being, specifically on 12 domains: affect, community, culture, education, economy, environment, health, human settlements, government, psychological/mental well-being, satisfaction with life and work | No | - | Positive and negative | No, only exemplarily; process-oriented |
| 2025 | Thomaidou and Limniotis [53] | All types of AI, particularly those utilizing personal data | Risks for data governance/protection and human rights | Yes | Distributional, recognition, and procedural justice | Negative | No; checklist format |
| 2025 | von Zahn et al. [48] | All types of AI with focus on machine learning | Fairness, explainability, and performance | Yes | Distributional justice | Negative | Yes |
| 2024 | Xia et al. [47] | All types of AI | Not specified. Mentioned: technical attributes are reliability, societal / ethical aspects like data quality and fairness, risk-oriented impacts and safety issues e.g. | Yes | Distributional justice | Negative | No; process-oriented |
| | | | unintended, hazardous behaviors or ethical vulnerabilities | | | | |
| 2026 | Zhang et al. [42] | AI in smart grid | Ethics and cybersecurity: Data accountability, security robustness, explainability and transparency, traceability. | Yes | Distributional and recognition justice | Negative | No; conceptual |
| 2021 | Zicari et al. [39] | All types of AI | 7 key factors of trustworthiness1 | Yes | Distributional, recognition, and procedural justice | Negative | No; process-oriented |

## 4. The Energy Justice Impact Assessment (EJIA) framework for AI

In the following, the Energy Justice Impact Assessment (EJIA) framework for multidimensional impact of AI is proposed mapping each quantifiable impact indicator onto potentially impacted stakeholders. It adds an energy justice layer to standard economic, environmental, and social indicators by (1) disaggregating impacts by stakeholder group and (2) including subjective justice perceptions alongside objective outcome measures. EJIA's contribution is descriptive rather than normative: it renders differential impacts across stakeholder groups visible and comparable, while the interpretation of whether a given pattern constitutes an energy justice concern remains a normative judgment for the practitioner applying EJIA, informed by baselines and context-specific definitions and benchmarks. We discuss consequences of this approach in Sections 5 and 6.

Section 4.1 traces how this indicator set is derived: we show how the AI lifecycle stages and the three justice tenets jointly can produce specific mechanisms of injustice, and how each mechanism can be operationalized as a quantifiable indicator.

### 4.1 From lifecycle mechanisms to measurable indicators

In Section 2.2, some examples of AI's potential impacts on energy justice were presented. These effects can be tied systematically to the AI lifecycle. Chen et al. [6] organize justice-relevant bias along three lifecycle stages: data collection, model development, and model deployment; they map these onto the three justice tenets (distributional, recognition, procedural). Thereby, they provide a diagnostic account of where injustice mechanisms can arise. Building on this diagnostic logic, we develop it into a measurement framework: the identified mechanisms inform the collection of quantifiable indicators, which are disaggregated by the stakeholder groups they affect, and assessed at defined phases against a counterfactual, so that observed effects can be attributed to the AI application itself. We further extend Chen et al.'s [6] three stages with a fourth, continuous model adaptation, to capture justice-relevant feedback loops that only emerge after deployment, and broaden the scope from algorithmic bias to the economic and environmental impacts of AI deployment as well. This approach is consistent with recent AI life cycle standards such as ISO/IEC 42005:2025 [31].

In Figure 2, each cell describes how the three types of injustices can arise at each life cycle stage. Take the exemplary mechanism during model deployment, where AI outputs interact with real-world conditions, producing differentiated outcomes: this could mean that a well-developed home energy recommendation system might give the same tip on thermostat settings to two different households leading to diverging outcomes, because both houses have different levels of insulation. This differential effect of AI on distributional justice could be uncovered by quantifying thermal comfort in all types of households the system is deployed to, and before and after deployment. “Thermal comfort” is therefore one outcome indicator which can help to quantify the effects of a newly developed and/or deployed AI system on different stakeholder groups.

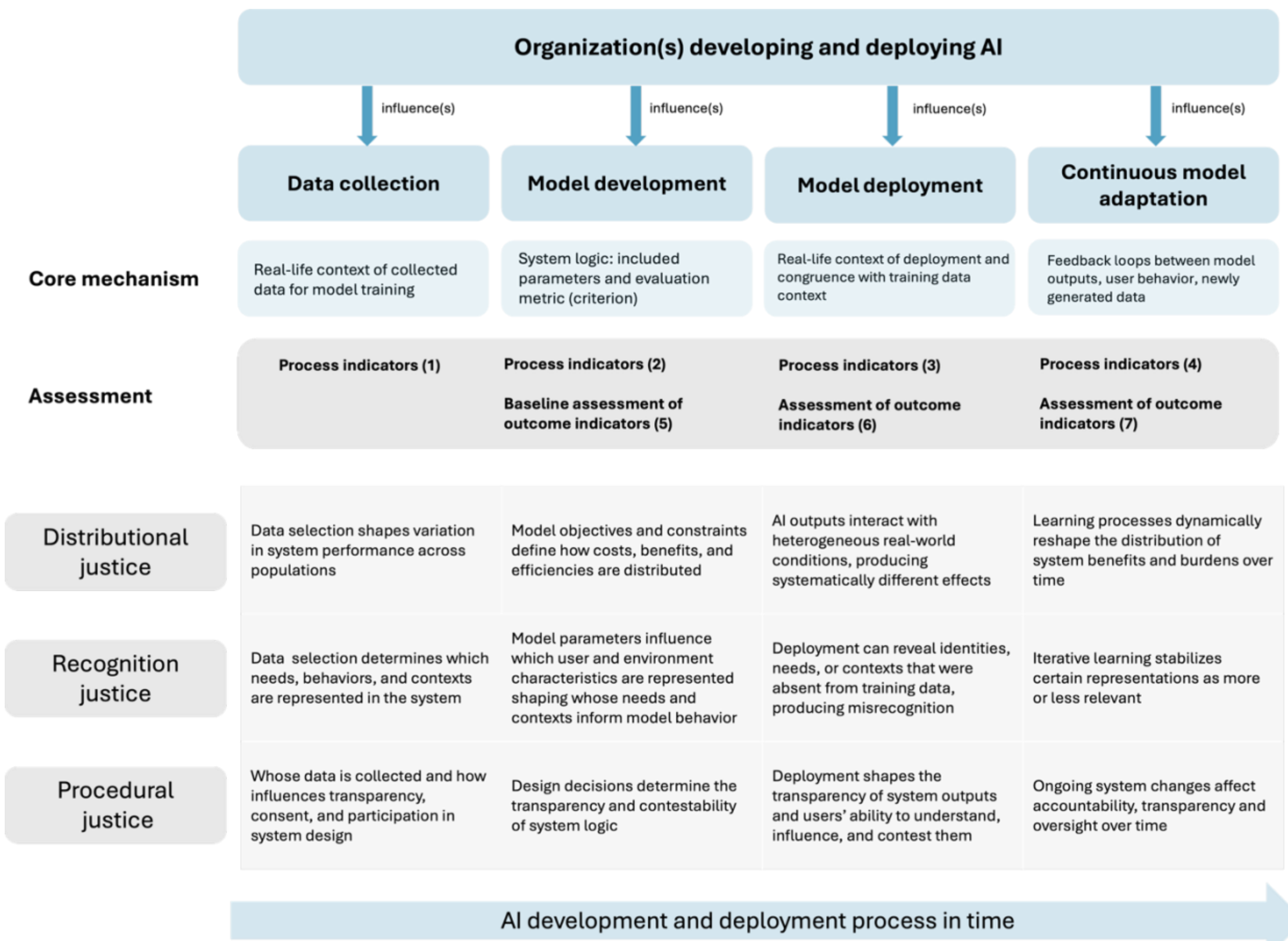

**Figure 2.** Mechanisms through which justice issues can occur derived from analyzing AI lifecycle phases in relation with each justice tenet. Phases are adapted from Chen et al. [6] and extended by the fourth phase continuous model adaptation.

In Figure 2, different assessment types and phases are categorized. Process indicators can be assessed at each stage of AI development and deployment and describe how decision-making is structured across the AI life cycle, focusing on who is involved, how decisions are taken, how responsibilities are assigned and how information is accessed, shared, and used. Process indicators primarily operationalize procedural justice, and, through indicators such as representativeness, partly recognition justice as well. They are complemented by outcome indicators, which allow to capture distributive and recognition impacts once the AI has produced observable real-world effects.

Outcome indicators need to be assessed relative to a control condition, as is the gold standard in experimental design [59]. There are three phases of assessment of outcome indicators: first, during model development, ensuring this assessment takes place before real-world deployment, second, when the AI has been deployed, and then third, after the model has been adapted. This latter assessment phase can be repeated until the outcome of the indicators is satisfactory, and therefore the model has shown to evoke the desired effects on the respective targeted stakeholder groups. This three-phased assessment approach allows for pre-post measurements of the selected outcome variables. Ideally, to be able to claim causality, each outcome variable should be measured in a control and an experimental condition during each assessment phase. This could mean assessing the same indicator in a real-world environment, for example a building that uses the AI tool that is to be assessed, and a comparable building that does not use this AI tool [60]. Then one can compute the difference between both conditions. This enables the assessment of changes attributable to the AI application, ideally through a difference-in-differences approach [61] or, where a control condition is not feasible, through longitudinal before-after comparisons with suitable model-side confound controlling, done across different stages of development, deployment, and adaptation.

### 4.2 Indicators to measure multidimensional impacts and their justice implications

Table 2, Table 3, and Table 4 demonstrate a collection of process and outcome indicators applicable to AI solutions in energy systems. For each AI system that is to be assessed, adequate indicators must be identified, as not all indicators will be applicable to all AI solutions and deployment contexts. For each indicator the assessment phases according to Figure 2 are indicated. Each indicator should optimally be assessed for each affected stakeholder group, though of course, some indicators may affect only one.

The selection of affected stakeholder groups is critical to the validity of the assessment. Rather than treating "stakeholders" as a single undifferentiated category, practitioners should disaggregate groups along dimensions that are plausibly linked to differential exposure to, or impact from, the AI application; this could mean distinguishing between groups defined by their structural relationship to the technology (for instance, residents of different housing complexes or building types, tenants versus building owners, or households on different tariff structures) where these distinctions correspond to meaningfully different exposure to the AI's decisions or outputs. Where relevant, disaggregation should also account for demographic characteristics that are known or hypothesized to correlate with differential vulnerability or benefit, such as income level, age, household composition, or energy poverty status. This is particularly important given that AI-driven optimization in energy systems (e.g., dynamic pricing, load-shifting incentives, or predictive maintenance prioritization) can generate distributional effects that are invisible at the aggregate level but significant for specific subgroups.

**Table 2**

Energy Justice Impact Assessment (EJIA) framework: environmental indicators.

| | Indicator | Description | Method/Metric | Assessment Phase (see Figure 2) |
|---|---|---|---|---|
| **1** | **Change in GHG Emissions** | Change in GHG emissions of AI training and due to use of AI | tCO2eq/ year | 5, 6, 7 |
| **2** | **Change in Energy Balance** | Change in net balance of energy during AI training and due to use of AI | kWh | 5, 6, 7 |

| | Indicator | Description | Method/Metric | Assessment Phase (see Figure 2) |
|---|---|---|---|---|
| **3** | **Change in Carbon Intensity** | Change in CO2 per unit of generated and/or used electricity | kgCO2/ kWh | 5, 6, 7 |
| **4** | **Change in Climate Adaptation** | Extent to which a system's ability to withstand and recover from climate-related stresses (e.g. extreme weather) is changed | Resilience score | 5, 6, 7 |
| **5** | **Change in Noise Pollution** | Change in environmental noise levels | dB | 5, 6, 7 |
| **6** | **Change in Renewable Energy Share** | Change in the proportion of energy demand met by renewable sources | **%** | 5, 6, 7 |
| **7** | **Change in Peak Demand** | Change in the highest (average) electricity demand observed during a defined period | kW/MW | 5, 6, 7 |
| **8** | **Change in Grid Integration & Stability** | Change in ability of the grid to handle variable supply and demand | Grid stability metrics | 5, 6, 7 |
| **9** | **Change in System Reliability** | Change in ability of the energy system to operate without unplanned interruptions due to AI-supported optimization (e.g., predictive maintenance) | % downtime | 5, 6, 7 |
| **10** | **Change in Material Resource Efficiency** | Change in amount of material required per unit of energy generated | kg material/ kWh | 5, 6, 7 |
| **11** | **Change in Water Resource Efficiency** | Change in annual volume of water used | $m^3$ water saved/year | 5, 6, 7 |
| **12** | **Change in Byproduct Utilization** | Change in share of generated byproducts that are reused or valorized instead of discarded | % byproducts used | 5, 6, 7 |
| **13** | **Change in Equipment Lifespan** | Change in operational lifetime of equipment | Years | 5, 6, 7 |
| **14** | **Change in Reuse & Recycling Rate** | Change in share of materials recovered for reuse or recycling relative to total material output | % materials recovered | 5, 6, 7 |

**Table 3**

Energy Justice Impact Assessment (EJIA) framework: social indicators.

| | Indicator | Description | Method/Metric | Assessment Phase (see Figure 2) |
|---|---|---|---|---|
| Process Indicators | | | | |
| **1** | **Stakeholder involvement** | Type of involvement of stakeholder groups in technology development | Technology provider/implementer survey | 1, 2, 3, 4 |
| **2** | **Representativeness** | Proportion of different stakeholders (defined by target population) participating in surveys, focus groups, co-creation processes etc. | Stakeholder survey demographic data | 1, 2, 3, 4 |
| **3** | **Procedural Justice Perceptions** | Subjective perceptions on procedural justice | Stakeholder survey based on organizational justice measurement by Colquitt [62] | 1, 2, 3, 4 |
| Outcome Indicators | | | | |
| **4** | **Distributional Justice Perceptions** | Subjective perceptions on distributional justice | Stakeholder survey based on organizational justice measurement by Colquitt [62] | 5, 6, 7 |
| **5** | **Recognition Justice Perceptions** | Subjective perceptions on recognition justice | Stakeholder survey: measure needs validation, see example in Supplementary Materials | 5, 6, 7 |
| **6** | **Change in Attitudes towards AI** | Change in sum of stakeholders' acceptance of involvement of AI, and perceptions of AI transparency and trust | Stakeholder survey: ATAI [63] | 5, 6, 7 |
| **7** | **Change in Energy Literacy** | Change in stakeholders' device energy literacy, action energy literacy, and financial energy literacy | Stakeholder survey or test: to be adapted based on van den Broek [64] | 5, 6, 7 |
| **8** | **Change in AI Literacy** | Change in stakeholders' subjective (survey) or objective (test) AI literacy | Stakeholder survey: MAILS [65] and/or test: AICOS [66] | 5, 6, 7 |
| **9** | **Change in (Pro-) Environmental Awareness/Attitude** | Change in stakeholders' energy-related awareness and attitudes | Stakeholder survey: adapted from Karlin et al. [67]; System usage data | 5, 6, 7 |
| **10** | **Change in Energy Poverty Risk** | Change in households' energy expenditure share, and self-reported inability to keep home adequately warm / arrears on utility bills adapted from EPAH methodology [68] | Stakeholder survey (household composition, energy expenditure, income, ability to keep home adequately warm, arrears on utility bills); | 5, 6, 7 |

| | Indicator | Description | Method/Metric | Assessment Phase (see Figure 2) |
|---|---|---|---|---|
| | | | benchmarked against national EPAH indicator thresholds | |
| **11** | **Change in Thermal Comfort/Stress** | Change in hours above thermal stress threshold; Thermal stress operationalized as hours above defined thermal stress threshold (objective and subjective) | Environmental monitoring data benchmarked with EN 16798-1 [69] or ASHRAE 55 [70] standard; Stakeholder survey: ASHRAE 7-point scale [70] | 5, 6, 7 |
| **12** | **Change in Perceived Agency in Energy Use** | Change in stakeholders' perceived ability to make decisions regarding their energy consumption and/or production | Stakeholder survey: to be adapted from GSE [71] | 5, 6, 7 |
| **13** | **Change in Overall Quality of Life** | Change in stakeholders' perception of their overall quality of life | Stakeholder survey: WHO-5 [72] | 5, 6, 7 |
| **14** | **Social Inclusion in Technology Access and Participation** | Extent to which different demographic groups are represented among users and have equitable access to the technology | System usage data; Technology provider/implementer survey: Computation of deviation of user demographics from target population | 5, 6, 7 |
| **15** | **Change in Number of Jobs** | Change in number of job positions needed/occupied | Technology implementer survey | 5, 6, 7 |
| **16** | **Change in Job Profiles** | Number of job positions with changing profile demands | Technology implementer survey | 5, 6, 7 |
| **17** | **Change in Habitability of Urban Environment** | Change in composite habitability score to be built including relevant indicators e.g. air quality, volume of waste/population, noise pollution, green space adapted from Mahmoudzadeh et al. [73] | Data from technology implementer Environmental monitoring data | 5, 6, 7 |
| **18** | **Change in Community Cohesion** | Change in perception of community cohesion | Stakeholder survey. adapted from Sampson et al. [74] | 5, 6, 7 |
| **19** | **(Change in) User-Friendliness of Technology** | Change in sum of survey-based e.g., user satisfaction, ease of use, and objectifiable data e.g., number of technology accesses or downloads | Stakeholder survey e.g. SUS [75]; System usage data | (5), 6, 7 |

| | Indicator | Description | Method/Metric | Assessment Phase (see Figure 2) |
|---|---|---|---|---|
| **20** | **(Change in) Effective Use of Technology** | Change in the extent to which stakeholders meaningfully engage with and benefit from the technology's outputs (e.g. recommendations followed; reduction in redundant or repeated system alerts) | System usage data | (5), 6, 7 |
| **21** | **AI safety** | Absence of typical AI-risks e.g. based on Slattery et al. [33] | Technology provider/implementer survey | 6, 7 |

**Table 4**

Energy Justice Impact Assessment (EJIA) framework: economic indicators.

| | Indicator | Description | Method/Metric | Assessment Phase (see Figure 2) |
|---|---|---|---|---|
| **1** | **Net financial impact** | Net financial impact over defined timeframe i.e. net costs or profit given investments, avoided costs and financial gains | System usage data; Technology implementer survey; Stakeholder survey | 5, 6, 7 |
| **2** | **Change in hours of working time** | Change in weekly working hours required over given timeframe via time-tracking before and after adoption of AI | Technology implementer survey; Stakeholder survey | 5, 6, 7 |
| **3** | **Years to Return on Investment (ROI)** | Time until cumulative benefits exceed technology investment costs | Financial data from implementing organization | 6, 7 |
| **4** | **Change in company performance** | Composite indicator capturing changes in company performance based on: Change in net profit margin (%), revenue growth (%), change in number of contracts signed (%) | Financial and operational data from implementing organization; components normalized and aggregated with equal weighting | 5, 6,7 |
| **5** | **Average amount customers are willing to pay** | Stated or revealed willingness to pay for added value by AI or new technological solution | Consumer research survey | (5), 6, 7 |

| | Indicator | Description | Method/Metric | Assessment Phase (see Figure 2) |
|---|---|---|---|---|
| **6** | **Changes to business model** | Impact of new technology on existing or requirement for new business model | Technology implementer survey | 6, 7 |
| **7** | **Scalability** | Share of technology components (%) reusable in other contexts without modification | Technology implementer survey | 6, 7 |

## 5. Exemplary application scenario of the EJIA framework

To illustrate how EJIA can be used to assess AI in energy, a brief application scenario in a social housing energy community in Portugal is presented here[2]. In this social housing complex, tenants have two different energy sources: Solar panels provide the tenants with free solar energy when available, and additionally, tenants are connected to the general grid, where they purchase power via individual contracts with commercial energy providers. An AI solution will optimize the tenants' energy consumption via scheduling recommendations, in order to help them lower their energy bill by allowing them to consume the most solar energy possible, e.g., put on the washing machine when the sun is out to maximize their solar energy consumption.

In order to quantify the impact of this AI-based application on distributional, recognition, and procedural justice, first, we identify relevant impact indicators and stakeholder groups (see Table 5) and design an assessment plan to schedule when to collect each indicator for which stakeholder group (see Table 6). For this exemplary scenario, we assume that the AI application was finalized in January 2026 and first applied in May 2026. Lastly, once all indicators have been assessed according to the assessment plan, effects can be visualized in a heat map (see Figure 3). An example of a possible survey that could be employed at the three assessment stages can be found in Supplementary Materials.

[2] As mentioned in the Acknowledgments, this publication has been prepared within the scope of the European Union Horizon Europe Innovation Actions project COSMIC. The social housing complex being the subject of this case study is part of the participating organization Porto Energy Agency, who are managing the energy concept of the municipality's social housing complex Agra do Amial.

**Table 5**

Exemplary definition of potentially affected stakeholder groups.

| Stakeholder group name | Group description |
|---|---|
| A | Tenants with disabilities living in social housing |
| B | Tenants without disabilities living in social housing |
| C | Social housing managers |
| D | Municipality owning the social housing |

**Table 6**

Exemplary plan for EJIA application.

| Indicator | March 2026 (pre-deployment) | June 2026 (post-deployment) | September 2026 (second iteration) |
|---|---|---|---|
| Environmental indicators | | | |
| Change in GHG emissions | A, B, D | A, B, D | A, B, D |
| Change in Energy Balance | A, B, D | A, B, D | A, B, D |
| Change in Renewable Energy Share | A, B, D | A, B, D | A, B, D |
| Social indicators | | | |
| Stakeholder Involvement | A, B, C, D | A, B, C, D | A, B, C, D |
| Change in Energy Literacy | A, B, C | A, B, C | A, B, C |
| Change in Energy Poverty Risk | A, B | A, B | A, B |
| Economic indicators | | | |
| Net financial impact | A, B, D | A, B, D | A, B, D |
| Change in hours of working time | C | C | C |
| Average amount customers are willing to pay | D | D | D |

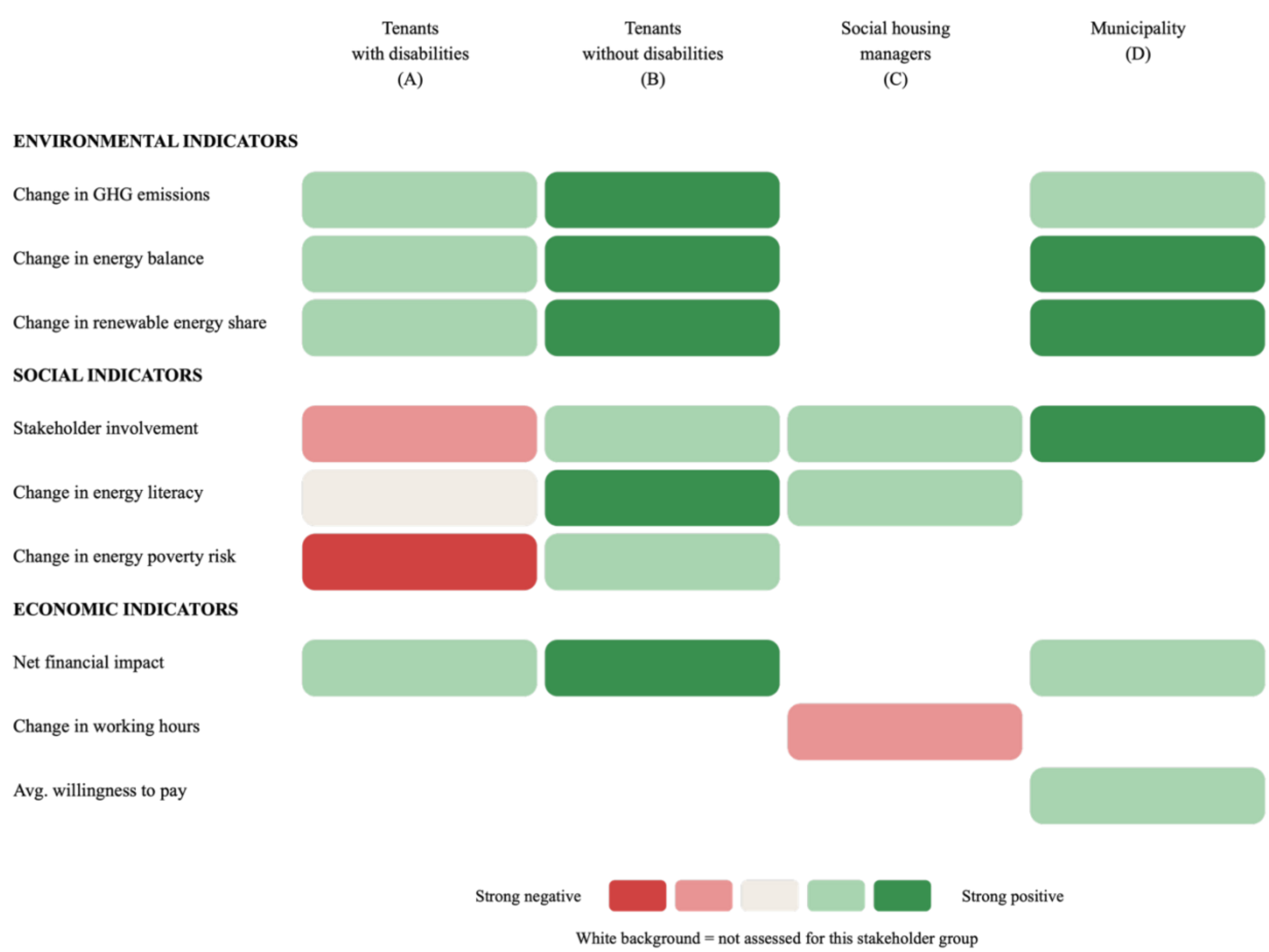

**Figure 3**. Exemplary heat map of EJIA applied to application scenario.

The heat map in Figure 3 shows that assessed impacts differ across stakeholder groups, with group A affected most negatively overall. While tenants without disabilities (B) show gains across social and economic indicators, tenants with disabilities (A) show no change in energy literacy, an increase in energy poverty risk, little improvement in net financial impact, and comparatively low stakeholder involvement during development.

As outlined in Section 4, descriptive differences do not by themselves establish a normative energy justice impact. Their interpretation requires case-specific context and requirements analysis that indicators alone cannot provide. Net financial impact illustrates this directly: a monthly saving of 10 euros for group A compared with 20 euros for group B could suggest that group B benefits more in absolute terms. Yet if group A's baseline energy expenditure is 30 euros per month and group B's is 100 euros, the 10 euros saving represents a substantially larger relative gain for group A (33 percent versus 20 percent). Conversely, group B may still experience the greater material benefit in absolute terms, which may matter for tenants' ability to meet their energy costs. Whether the disparity constitutes an energy justice concern therefore depends on the relevant baseline, the distributional objective, and the material consequences for each group. Suppose, however, that after accounting for the baseline, group A's outcome is still judged to constitute an energy justice concern. Even then, the appropriate response depends on why the disparity exists, as different causes point to different interventions. The disparity could arise from the AI system's design, for instance if the scheduling recommendations are poorly suited to group A's daily routines or mobility constraints, calling for model recalibration. Or it could result from implementation, i.e. unequal onboarding support or structural barriers such as appliance ownership or contractual arrangements that limit group A's ability to act on the recommendations, calling for changes to the rollout rather than the model itself. Ultimately, indicators can provide the evidence base for identifying disparities; contextual information (in the above, information on group A's mobility constraints, or onboarding records) is needed to interpret their justice implications and determine an appropriate response.

## 6. Discussion

Even though AI has widespread applications both on the production and the consumption side in the energy transition [2,12,13], little empirical research has been examining AI's impact on energy justice. Existing studies are mostly qualitative or simulation based [e.g., 6,18,19]. Therefore, little is known today about real-world positive or negative impact of AI in energy. Further, to date, no quantitative framework exists allowing assessment of economic, environmental, and social impacts of AI on different stakeholder groups, much less in the context of energy systems as established by Stahl et al. [37] and by the systematic literature review in Section 3. Most existing impact assessment frameworks are compliance [e.g., 29,30] or solely process-oriented, without complementary outcome indicators capturing distributive effects [e.g., 45–47]. Addressing this gap requires a systematic way of deriving quantifiable indicators: crossing the stages of the AI lifecycle with the tenets of energy justice [6] surfaces the specific mechanisms through which injustice can arise.

In this paper, we therefore propose the EJIA framework for AI. EJIA allows to quantify both positive and negative impact of AI, which can then be compared with each other across different stakeholder groups and indicators. This can be done by first identifying relevant stakeholder groups and secondly selecting relevant indicators for each stakeholder group. Through the implementation of a measurement plan, indicators are then measured across time to be able to attribute indicator outcomes to the implementation of AI. Economic, environmental, and social impact can be assessed at the same time allowing to get a balanced overview of impact on different dimensions. By measuring indicators for those different dimensions across different stakeholder groups separately, differential impacts relevant to energy justice become visible, providing the evidence base from which justice implications can be judged. As set out in Section 4.1, process indicators primarily operationalize procedural justice, while outcome indicators capture distributive and recognition impacts once the AI has produced real-world effects. Stakeholders' self-reported justice perceptions add a subjective layer across both.

The presented list of indicators is non-exhaustive but covers a broad range of application contexts, as it was developed within the scope of the EU Horizon project COSMIC in workshops with stakeholders across many pilot sites and use cases. They reflect relevant use cases such as DSR systems, predictive maintenance, or AI-powered housing development, but, as real-world application possibilities for AI in energy are numerous, so is the universe of possible corresponding impact indicators. EJIA can therefore be used both as a toolbox to select adequate indicators out of the proposed collection, but also as a methodology to additionally develop use case specific indicators.

Even though the goal of EJIA is the quantification of objectively measured impact, by choosing relevant stakeholder groups and indicators, practitioners can in part determine the overall outcome of their impact assessment. To mitigate this, we recommend that the assessment plan, including the stakeholder groups and indicators selected as well as those deliberately excluded, be documented and made transparent prior to measurement, and that, ideally, affected stakeholder groups be involved in the selection process itself. Reporting all assessed indicators, including those showing no or adverse effects, provides a full evidentiary basis for interpreting the assessment, showcases trade-offs, and increases transparency.

Further, the main contribution of the framework is to create a basis for technology adaptation by making multidimensional impact visible. It does not, however, prescribe how different indicators should be weighted or aggregated. This is a deliberate feature: whether an observed disparity constitutes an energy justice concern depends on the baseline against which it is assessed, the stakeholders considered, and the distributive principle applied. The same pattern of outcomes may support different justice judgments. An assessment based on equality of absolute outcomes may reach a different conclusion from one emphasizing proportional change, minimum thresholds of sufficiency, or the needs of particularly vulnerable groups. These normative choices cannot be derived from the indicators themselves and must instead be made explicit and justified in the context in which the framework is applied.

The same applies to decisions about technology adaptation. Identifying a disparity in an indicator does not in itself indicate whether the AI system, its implementation, or the wider conditions in which it operates should be changed. Diagnosing the source of an observed impact requires contextual evidence beyond the indicator values, such as information on system design, implementation processes, affected groups, and relevant socioeconomic conditions. EJIA therefore does not prescribe a single definition of a just outcome, nor which adaptation follows from a given result. Rather, it provides an evidence base for making multidimensional impacts visible and supports context-sensitive decisions about whether and how an AI system should be adapted.

## 7. Conclusion

This paper introduced the Energy Justice Impact Assessment (EJIA) framework for AI, developed in response to a gap identified through a systematic review of 26 existing impact assessment approaches: we found no peer-reviewed framework that combines multidimensional impact coverage with quantifiable indicators that can be disaggregated by stakeholder group. By crossing the stages of the AI lifecycle with the tenets of energy justice, the framework provides a structured basis for deriving environmental, social, and economic indicators, for measuring them across process and outcome phases, and for comparing effects between the groups an AI application affects. The exemplary application scenario illustrates how the resulting evidence can inform decisions about technology adaptation rather than remaining descriptive.

## Acknowledgments

Funding: This work was supported by the European Union's Horizon Europe Innovation Actions programme under grant agreement No 101189676 (COSMIC).

## Declaration of generative AI and AI-assisted technologies in the manuscript preparation process

During the preparation of this work the authors used Claude AI in order to assist with copyediting and language refinement. After using this tool/service, the authors reviewed and edited the content as needed and take full responsibility for the content of the published article.

## Author contributions: CRediT

**Emily Bringmann**: Conceptualization, Methodology, Investigation, Formal Analysis, Visualization, Writing – Original Draft, Writing – Review and Editing. **Florian Kutzner**: Conceptualization, Methodology, Supervision, Writing – Review and Editing. **Celina Kacperski**: Conceptualization, Methodology, Supervision, Writing – Review and Editing.